\documentclass[conference]{IEEEtran}
\IEEEoverridecommandlockouts
\usepackage{cite}
\usepackage{amsmath,amssymb,amsfonts}
\usepackage{algorithmic}
\usepackage{graphicx}
\usepackage{textcomp}
\usepackage{hyperref}
\usepackage{xcolor}
\def\BibTeX{{\rm B\kern-.05em{\sc i\kern-.025em b}\kern-.08em
    T\kern-.1667em\lower.7ex\hbox{E}\kern-.125emX}}
\begin{document}

\title{Adaptive re-calibration of channel-wise features  for Adversarial Audio
Classification
}

\author{\IEEEauthorblockN{$Vardhan\ Dongre^{[1]}$, $Abhinav\ Reddy\ Thimma^{[1]}$, $Nikhitha\ Reddeddy^{[1]}$}
\IEEEauthorblockN{{\small[1]} University of Illinois Urbana-Champaign}
}
\maketitle

\begin{abstract}
DeepFake Audio, unlike DeepFake images and videos, has been a less studied topic, and the solutions which exist for the synthetic speech classification either use complex networks or don’t generalize to different kinds of synthetic speech. Through this work, we perform a comparative analysis of different proposed models for synthetic speech detection including End2End and ResNet-based models against synthetic speech generated using Text to Speech and Vocoder systems like WaveNet \cite{DBLP:journals/corr/OordDZSVGKSK16}, WaveRNN \cite{DBLP:journals/corr/abs-1802-08435}, Tactotron, and WaveGlow. We also experimented with Squeeze Excitation (SE) blocks in our ResNet models and found that the combination was able to get better performance. In addition to the analysis, we propose a combination of Linear frequency cepstral coefficients (LFCC) and Mel Frequency cepstral coefficients (MFCC) using the attentional feature fusion technique to create better input features which can help even simpler models generalize well on synthetic speech classification tasks. Our best models (ResNet based using feature fusion) were trained on Fake or Real (FoR) dataset and were able to achieve 95\% test accuracy with the FoR data, and an average of 90\% accuracy with samples we generated using different generative models.
\end{abstract}

\begin{figure*}[htp]
    \centering
    \includegraphics[width=15cm]{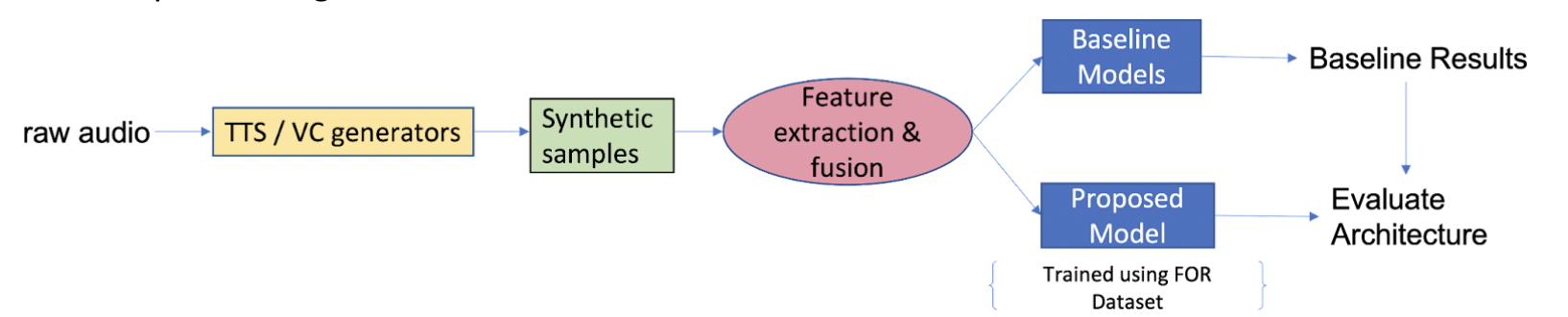}
    \caption{Workflow Pipeline}
    \label{fig:model_workflow}
\end{figure*}

\section{\textbf{Introduction}}
Speech synthesis and spoofing attacks have become prevalent in recent years because of the development of generative models which are able to synthesize speech of great quality which humans are not able to  distinguish. The spoofing attacks include either replay attacks, where the speaker's voice is recorded and used in a different context, or generated speech attacks, where a text to speech or voice conversion system is able to generate new voice samples. In this work, we focus on the generated speech attacks from both text to speech and voice conversion systems because this is an area that is rapidly changing because of newer neural network-based generative models. We perform a comparison of performance between different existing models and proposed models, against a variety of generated speech samples created using WaveNet \cite{DBLP:journals/corr/OordDZSVGKSK16}, WaveRNN \cite{DBLP:journals/corr/abs-1802-08435}, Tacotron\&WaveGlow \cite{DBLP:journals/corr/WangSSWWJYXCBLA17}, and FastSpeech \cite{DBLP:journals/corr/abs-1905-09263}. 
\par{The speaker verification community has been able to come up with innovative models to tackle the problem described above. A majority of models in the synthetic speech detection domain fall under two categories: Traditional models and End to End systems [Fig \ref{fig:model_workflow}]. Traditional systems try to tackle the problem in two phases. The first of which is feature extraction and the next is building a classifier based on the extracted features. End to End systems skip over the feature extraction phase and build models which take in the raw audio samples as input and give out a classification result. Traditional models with specifically curated features have been shown to produce promising results in this domain, but the newer End to End systems are not that far behind and are to get similar performance without much focus on features. End to End systems have been a recent solution for the problem and had issues with noisy audio during our initial analysis. Another major issue with the existing systems is the generalizing capability to speech synthesized using newer generative models. Thus we went ahead with a traditional architecture using ResNets and explored the impact of feature fusion on features extracted from speech.} 

\par {For Audio classification tasks features like Mel-Spectrogram, MFCC, LFCC, CQT and CQCC have been to produce good results. However, LFCC and MFCC are the features we focused on specifically for building the classifier. MFCC has been used in a variety of speech recognition applications and has been promising in detecting the lower frequencies coefficients from human speech. LFCC on the other hand was added so as to introduce a measure for higher frequency coefficients which might be prevalent in the synthesized speech. We focused on fusing both LFCC and MFCC and settled on using Attentional Feature Fusion (AFF) technique \cite{dai2021attentional}. AFF has shown promising results in fusing and scaling features of inconsistent semantics. We built a total of 8 models using 3 combinations of features: only LFCC, only MFCC, and a combination of LFCC and MFCC. In these 8 models, we experimented with 2 different ResNet structures: ResNet34 and ResNet50, and compared the results with pre-trained versions of ResNets. We also introduced Squeeze excitation blocks into our model architecture and found that it was able to improve the performance of our models by enhancing the inter-channel dependencies for our binary classification task.}

\par{For the problem, we used the Fake or Real normalized dataset which has an even distribution of samples between genders (male and female) and classes (fake and natural). This dataset was primarily used for training our models. We evaluated the trained models on the Fake or Real datasets' test samples and also against all of the generated audio samples. To augment the test set, we also verified the performance of our models against the ASVSpoof 2019 dataset \cite{wang2020asvspoof}. } 
\par{Our primary contribution through this work is the introduction attentional feature fusion block, which has been effective in the image domain,  into the speech domain to leverage the combination of different extracted features. The paper is organized as follows: Section 2 discusses the background of feature engineering, feature fusion, and existing model architecture. Section 3 focuses on our experimentation setup, baseline, and proposed models. In Section 4, we discuss the results from our testing and compare different models we experimented with. The conclusion of the paper would be in Section 5.}

The code for this work can be found at \url{https://github-dev.cs.illinois.edu/athimma2/deepfake-audio-classifier}

\begin{figure}[htp]
    \centering
    \includegraphics[width=5cm]{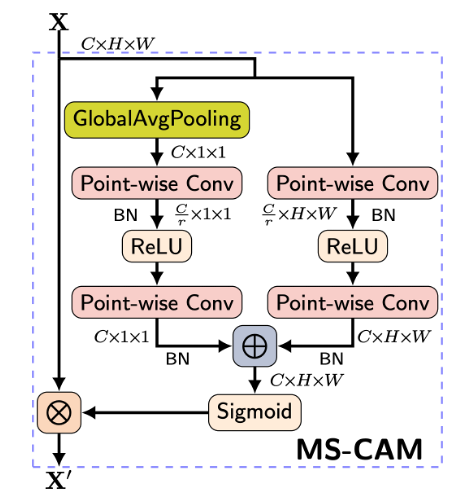}
    \caption{A Multi scale channel attention module used inside AFF}
    \label{fig:mscam}
\end{figure}

\begin{figure}[htp]
    \centering
    \includegraphics[width=5cm]{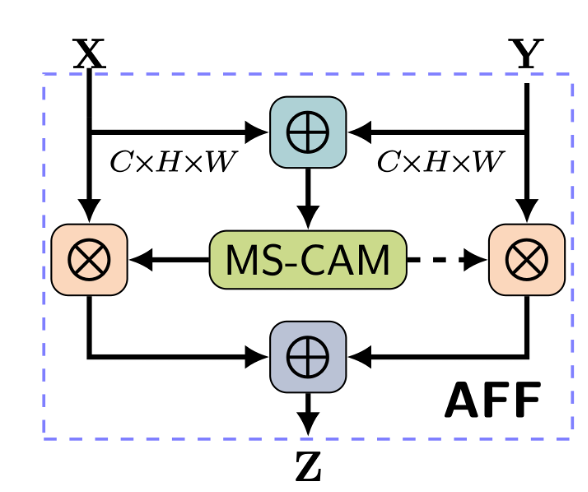}
    \caption{The Attentional feature fusion block}
    \label{fig:aff}
\end{figure}

\section{\textbf{Background}}

\subsection{\textbf{Feature Engineering}}

Feature engineering is an essential component of learning algorithms, the performance of the ML models is heavily dependent on how we represent the feature vector. As a result, a significant time and effort is spent in designing preprocessing pipelines and data transformations. In the audio domain, audio files usually exist in the form of digital files with wav, .mp3, .wma, .aac, .flac etc. as the common formats. The major audio features extracted from them can either be timbral texture features or rhythmic content features. In this work our focus has been only on using timbral texture features which specifically include MFCC and LFCC. A common practice while modelling deep learning frameworks in the audio domain is to convert the audio into spectrogram which is a concise snap of an audio wave that has undergone a Fourier transform. Mel-frequency cepstral coefficients (MFCC) is a cepstral representation of the audio which has been widely used in automatic speaker recognition and vocoder systems. Introduced in the 1980’s, they have been the state of the art ever since as they have proven to be robust in training several deep learning algorithms for high level audio domain tasks. Linear-frequency cepstral coefﬁcients (LFCC) is another alternative to MFCC which has been also used as a go to feature for training models to obtain learnable parameters, the difference between LFCC and MFCC is based on the filter banks that they use for transforming the audio, where MFCC uses a Mel filter bank, LFCC uses a linear filter bank. Several studies have shown both features to be comparable although MFCC is dominantly used in speakers as well speech recognition still. 

\subsection{\textbf{Feature Fusion}}

Many times, a single feature representation obtained from the data is insufficient to convey the necessary details of the underlying distribution of the natural process, thus a common approach in many feature engineering methods is fusing together features obtained from the single data source through different methods. This combination of features can either be a simple concatenation or can be an informed mathematical combination function.  In the image domain fusing features obtained from different layers is not a novel idea, image pyramids are one such example of this attempt. However, while combining such cross-layer features a common problem we run into is of scale variance. In our framework we decided not to rely on simple concatenation or summation of features but rather include scale invariant fusion mechanism that has been recently introduced in the image domain. Attentional feature fusion (AFF) \cite{dai2021attentional} [Fig \ref{fig:aff}] is a trainable mechanism that has the ability to fuse together features obtained across long and short skip connections without running into issues of scale variance. The attention module MS-CAM [Fig \ref{fig:mscam}] in AFF aggregates the contextual information along the channel dimension from different receptive fields which makes the fusion process more dynamic and informed.

\subsection{\textbf{Squeeze Excitation}}
\begin{figure}[htp]
    \centering
    \includegraphics[width=7cm]{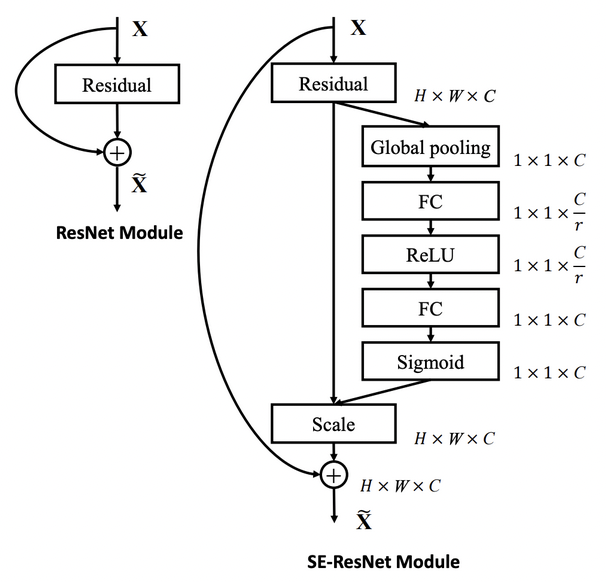}
    \caption{Squeeze Excitation block with ResNet architecture}
    \label{fig:SE_block}
\end{figure}

A common theme across deep learning research has been obtaining powerful representations from the data that capture specific properties that are most useful for the task at hand. Squeeze Excitation block \cite{hu2018squeeze} [Fig \ref{fig:SE_block}] is an architectural unit that has been introduced specifically for this purpose. But instead of combining spatial correlations the SE block focuses on modeling the interdependencies across the channels of the features obtained from CNNs. As a result of such modelling and re-calibration, the more useful features are emphasized, and less useful ones get suppressed. From Fig we can see the structure of the block which can be understood in two separate operations viz., squeeze operation and excitation operation. The squeeze operation as the name suggests collects and squeezes together all the feature maps to generate a descriptor. The excitation operation then takes this descriptor embedding to generate weights for each channel. Once we obtain the weights, they are applied to the input feature map to generate the output for this block.

\subsection{\textbf{Existing Approaches}}
From extensive literature study on the problem of identifying spoofed audio and classification of deep fakes, we came to the conclusion that most of the existing approaches in this area either rely on classical ML algorithms that depend on probabilistic modeling or exploiting the temporal information of the audios or there existed a wide range of complex deep learning models with millions of parameters consisting of gating operations which were significantly bulkier. Also due to the rapid advances in the generative algorithms that can create indiscernible synthetic audios the performance of these methods needed to be tested on these novel synthesizing algorithms.
\par{One major baseline for our work was the End2End model \cite{hua2021towards} [Fig \ref{fig:baseline_model}] specifically the Time-domain Synthetic Speech Detection Net (TSSDNet) based on a Inception CNN. The underlying theme of this work is to show that there is no need for any hand crafted features for detection of synthetic speech; rather a model could be trained directly on raw audio inputs to classify the natural samples from synthetic ones. The Inception style TSSDNet solely relies on identifying the forensic artifacts formed due to the synthesizing process. Since such artifacts have a greater chance of being observed in the features obtained in the upper layers the model therefore the model is shallower. This work tested its framework on the ASVspoof dataset \cite{wang2020asvspoof} and demonstrated that the TSSDNet significantly reduced the cross dataset Equal Error Rate. Another thrust in this problem was through the introduction of Res2Net architecture \cite{li2021replay} as shown in the which utilizes a residual connection across different channel groups split from the input feature map. The underlying of this approach lies within creating multiple scaled feature fields. This work experimented with several features derived from the ASV dataset \cite{wang2020asvspoof} but found CQT as the most informative feature set.}

\section{\textbf{Experimental Setup}}
\subsection{\textbf{DataSets}}
In this section, we describe two types of datasets. FoR dataset \cite{reimao2019dataset} that is used for training and testing phases and ASVSpoof2019 Logical Access (LA) dataset \cite{wang2020asvspoof} to evaluate the effectiveness and generalization power of fake speech detection models. We later introduce the metrics to evaluate the performance.

\subsubsection{\textbf{FoR-norm Dataset}}
The Fake or Real (FoR) dataset \cite{reimao2019dataset} is proposed for studies in speech synthesis and synthetic speech detection. The FoR database contains speech clips from state-of-the-art speech synthesis algorithms, i.e. speech clips with naturalness similar to real human speech. The fake speeches are synthesized by the latest methodologies, both open source and commercial, such as: DeepVoice, Amazon AWS Polly, Baidu TTS, Google TTS, and Microsoft Azure TTS. In addition, the FoR dataset contains a large number of speech clips that are enough to train complex models without overfitting. \newline
\par{To eliminate bias for machine learning experiments, the FoR database is processed into four different versions: FoR-original, FoR-norm, FoR-2seconds, and FoR-rerecorded. The FoR-norm dataset is processed by balancing the data to achieve an even distribution between genders (male and female) and classes (fake and natural). Due to the balancing process, the FoR-norm version of FoR dataset \cite{reimao2019dataset} contains a total of 69,400 speech clips. To eliminate strange and unusual speeches, we use the FoR-norm version of the FoR database. The entire FoR-norm dataset is partitioned into training and testing sets. More details about the speeches in the FoR-norm dataset are shown in [Table-1] .}
\par{The main reason why we used the FoR dataset  \cite{reimao2019dataset} is the systems that this dataset used to generate fake speech samples are different from the ones that we used to collect the test data (described in the next section). So, if our models were able to correctly predict the true labels of these collected samples too, then we can say that our models are generalized and can differentiate samples from most of the speech synthesizers.}

\begin{table}[htbp]
\caption{Illustration of subsets and speeches in FoR-norm dataset.}
\begin{center}
\begin{tabular}{|c|c|c|}
\hline
\textbf{Subsets} & \textbf{\textit{Real Speech}}& \textbf{\textit{Fake Speech}} \\
\hline
Training & 16000 & 16000  \\
\hline
Testing & 4000 &4000 \\
\hline
\end{tabular}
\label{tab1}
\end{center}
\end{table}

\subsubsection{\textbf{Synthetic Speech Dataset Collection}}
To evaluate the generalization power of our models, we generated some fake speech samples that are collected using different Text-to-speech and Vocoder systems. We used four such systems 1. Wavenet \cite{DBLP:journals/corr/OordDZSVGKSK16} 2. WaveRNN \cite{DBLP:journals/corr/abs-1802-08435} 3. FastSpeech \cite{DBLP:journals/corr/abs-1905-09263} 4. WaveGlow \& Tacotron \cite{DBLP:journals/corr/WangSSWWJYXCBLA17} and generated fake samples which are later used in the evaluation phase of the project. \newline

\subsubsection{\textbf{Cross-Dataset Evaluation}}
{Apart from generating the fake speech samples from different sources, we also collected samples from another dataset. Given two labeled datasets that target fake speech detection, cross-dataset evaluations also aim to detect the fake speeches that are generated by completely unseen fake speech technologies. To explore how well the detection model trained on one dataset generalizes to another dataset, we trained the models on the FoR dataset \cite{reimao2019dataset} and evaluated the performance on the ASVSpoof2019 LA dataset. \cite{wang2020asvspoof}}

To summarize, the training and testing stages use only FoR norm dataset. The Evaluation stage is performed on 80 samples collected from LA dataset and from different speech synthesizers.

\subsection{\textbf{Front-end Feature Engineering}}
\begin{figure}[htp]
    \centering
    \includegraphics[width=8cm]{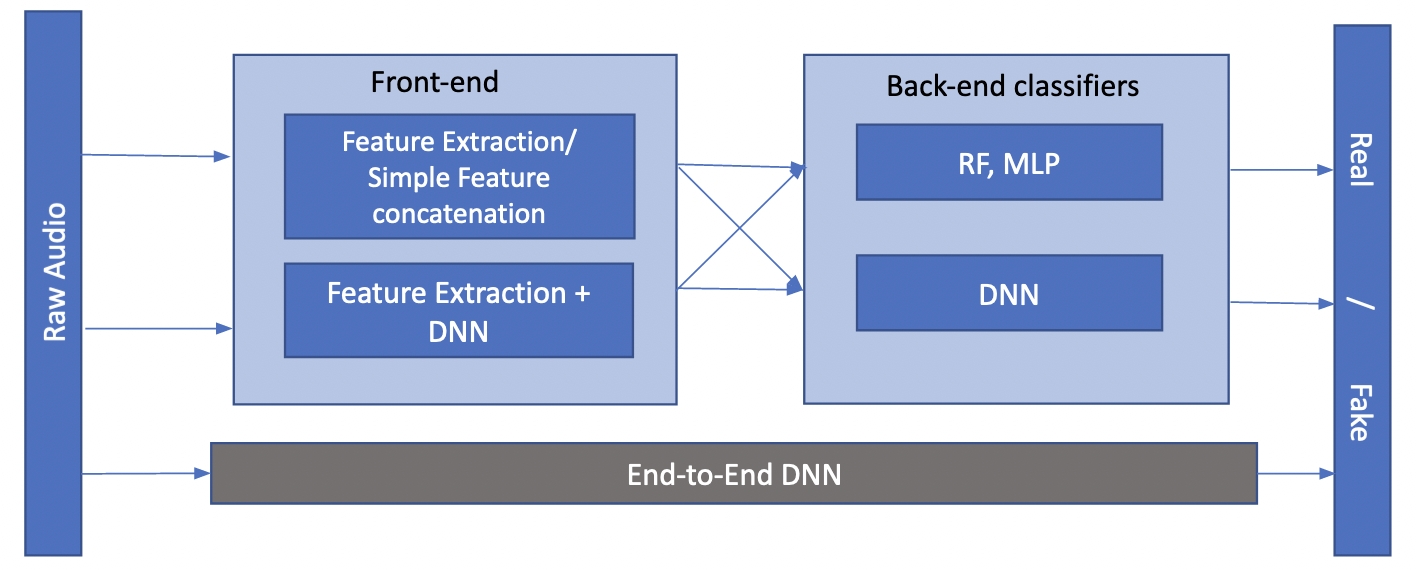}
    \caption{Existing Front-end/Back-end pipeline and End-to-end pipeline}
    \label{fig:baseline_model}
\end{figure}

The proposed models in our project follow the traditional architecture which is front-end feature engineering and back-end classification [Fig  \ref{fig:baseline_model}]. The front-end acoustic features are pretty important for efficiently detecting fake speech. Two simple and commonly used acoustic features 1. Linear frequency cepstral coefficients (LFCC) and 2. Mel frequency cepstral coefficients (MFCC) are evaluated in our experiments. All raw audio files are converted into multiple 1 sec samples and for each second we generate MFCC frames (20 x 44) hop-length = 512, and FFT = 2048 using librosa library and LFCC frames (136 x 13) window-hop = 0.01, FFT = 512 using spafe library. \newline
\par{LFCC and MFCC are largely identical in terms of coefficient extraction scheme. The only difference is in the filter bank, since the LFCC filter bank coefficients equally cover all speech frequency ranges and consider them of equal importance while MFCC considers only low-frequency ranges. So, the advantage of using LFCC is it improves the distinguishability of speaker-specific characteristics available in the higher frequency zone. How we use this advantage in training our features is described in the next section.}
 
\subsection{\textbf{Feature Fusion Strategy}}
The acoustic features generated above are fused together and are passed through our classifiers for prediction. For fusion, we have used AFF (Attentional feature Fusion) technique \cite{dai2021attentional}. To pass through the AFF, the features must be of the same size but the above obtained MFCC and LFCC samples are of different sizes. MFCC - (20 x 44) and LFCC (136 x 13). So, we padded both sample types with zeros on both ends and created a unique sample size of (136 x 44). These padded MFCC and LFCC data are passed through a single 2D Convoluted layer with a kernel of size 2x2, stride = 1 to extract features from them. After obtaining the convoluted features of both MFCC and LFCC, we fuse them by passing through the AFF block \ref{fig:aff} that is described in the background section.

\subsection{\textbf{Back-end Classification}}
\paragraph{\textbf{Baseline Models}}
\par{
As discussed in the background, we have different architectures for fake speech detection. There are traditional approaches that follow front-end feature engineering and back-end classification and a modern approach End-to-End neural network classification.
To define a starting accuracy that we want to improve further, we have used classical machine learning algorithms RandomForest and Multi-layer Perceptrons as our baseline models. We also compared our models’ performances with the existing End-to-End model \cite{DBLP:journals/corr/WangSSWWJYXCBLA17} to see if we achieved better performance when compared to this inception network model.} 

\paragraph{\textbf{Random Forest:}}
{A random forest is a meta estimator that fits a number of decision tree classifiers on various sub-samples of the dataset and uses averaging to improve the predictive accuracy and control over-fitting. We have formed 100 decision trees with a bootstrapping strategy to form subsets of samples.} 

\paragraph{\textbf{Multi-layer Perceptron:}}
{Multi-layer Perceptron (MLP) is a supervised learning algorithm that learns a non-linear approximator function $f : \mathbb{R}^{m} \rightarrow \mathbb{R}^{o}$  by training on a dataset, where m is the number of dimensions for input and o is the number of dimensions for output. We have used stochastic gradient descent optimizer to optimize the log-loss function. There are 2 hidden layers in our model, each with 5 and 2 hidden units respectively.}

{The above two baseline classifiers are trained and tested using individual MFCC and LFCC features that are obtained from the same subsets of the dataset defined in Table-1.} \newline

\paragraph{\textbf{End-to-End model:}}
The end-to-end model defined in \cite{hua2021towards} is named Time-domain Synthetic Speech Detection Net (TSSDNet) since it uses raw-audio sound waves in the time domain. In this TSSDNet, an advanced CNN structure is considered that includes Inception-style parallel convolution networks and the model is called Inc-TSSDNet. We obtained a pre-trained TSSDNet from [add reference] and tested it on the evaluation subset from Table-1. We didn’t achieve better results with this better since this was trained using the ASVSpoof2019 LA dataset \cite{wang2020asvspoof}. Therefore, the end-to-end model was not able to generalize and hence failed in predicting the samples from different synthesizers.\\

The comprehensive results of all the baseline models are detailed in the Results section.

\paragraph{\textbf{Proposed Models}}
\par{
We formed a total of 8 models using the combinations of Sequential Excitation blocks, ResNet blocks, and front-end acoustic features.}\\

\paragraph{\textbf{Training Strategy:}} We first experimented with following 4 models and developed 4 more models in order to fix the issues we observed with the first 4 models.
\begin{itemize}
\item SEResNet50 + MFCC
\item SEResNet50 + LFCC
\item SEResNet50 + AFF
\item SEResNet34 + AFF
\end{itemize}

\begin{figure*}[htp]
    \centering
    \includegraphics[width=15cm]{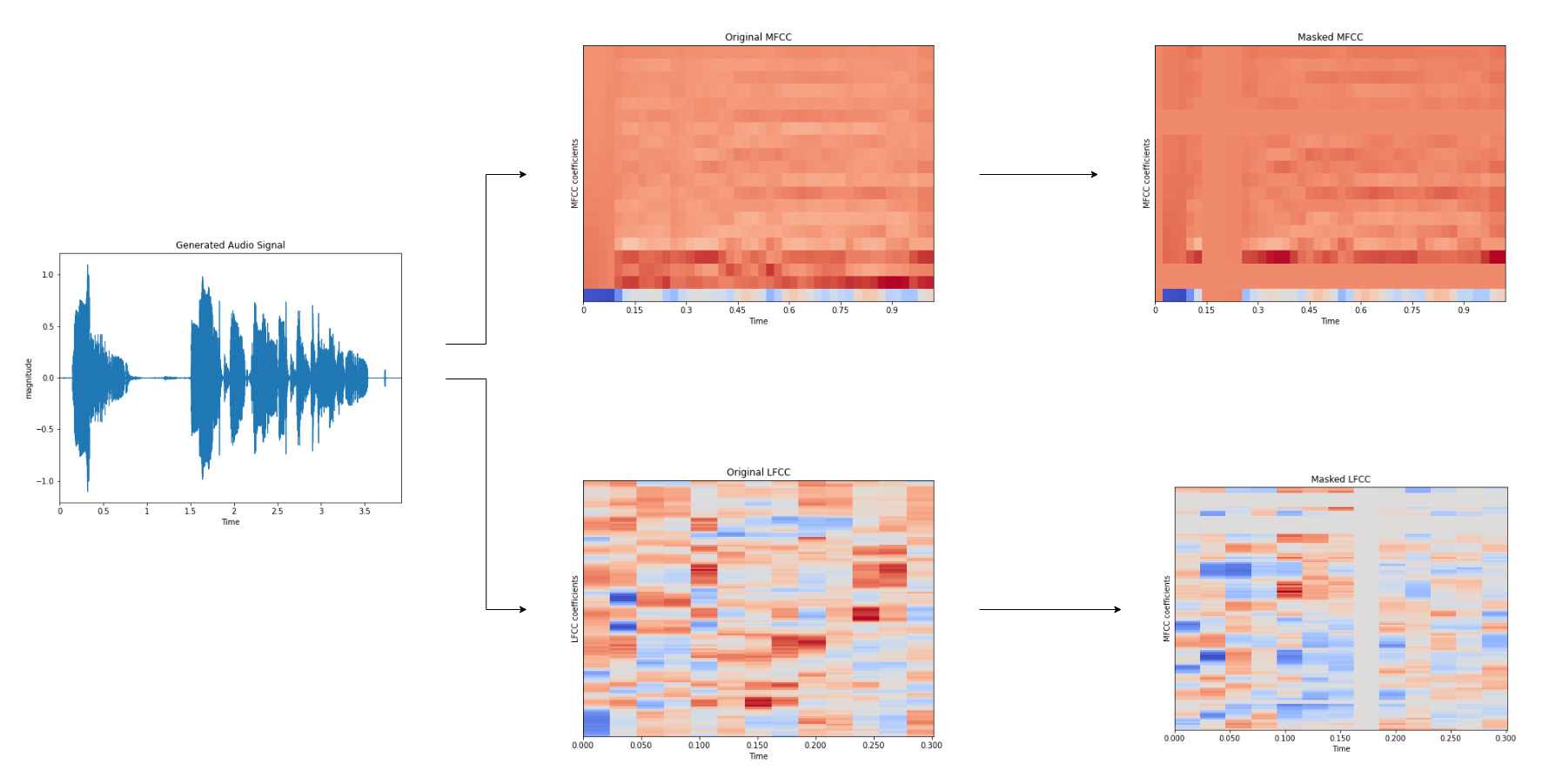}
    \caption{Image depicting the conversion of raw audio into LFCC and MFCC and masking the resultant features}
    \label{fig:Masking-image.jpg}
\end{figure*}

The first 2 models use the individual features extracted as the input to the SEResNet50 neural network. The Third model has the input AFF which means the fused features of MFCC and LFCC using Attentional Feature Fusion to the same SEResNet50 model. While training this model, we observed that the model was overfitting and failed to perform on the test data. This means that the model was not able to generalize the features of different synthesizers. This was also the case with the SEResNet34 model. To resolve this issue, we transformed the dataset by masking samples [Fig \ref{fig:Masking-image.jpg}]. 

\begin{figure}[htp]
    \centering
    \includegraphics[width=7cm]{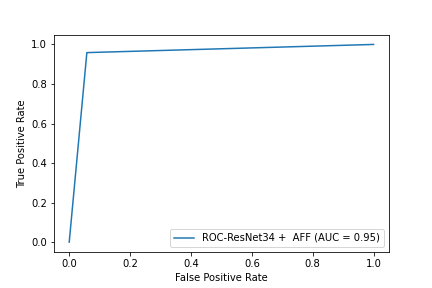}
    \caption{ROC curve for SE ResNet34 + AFF model}
    \label{fig:ROC_resnet34_AFF}
\end{figure}

\begin{figure}[htp]
    \centering
    \includegraphics[width=7cm]{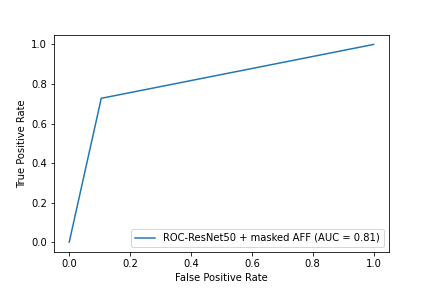}
    \caption{ROC curve for SE ResNet50 + masked AFF model}
    \label{fig:ROC_resnet50_masked}
\end{figure}

\paragraph{\textbf{Data Transformation:}} 
The data samples are transformed by creating a masked data where we used time masking and frequency masking techniques [Fig \ref{fig:Masking-image.jpg}]. These techniques select some range of randomly chosen frequencies over a single time frame and some range of time frames for a single frequency and all the selected bits are replaced with their respective mean values. This range is selected as 7\% in our models. By this process, we are hiding few characteristics of the features and generalize our original data. Hence by this data transformation, we created 4 more models which improved the accuracy. 
\begin{itemize}
\item SEResNet50 + Masked-MFCC
\item SEResNet50 + Masked-LFCC
\item SEResNet50 + Masked-AFF
\item SEResNet34 + Masked-AFF
\end{itemize}
Apart from the aforementioned 8 models, we also used a pure ResNet model (without SE block) that is trained on image dataset to see its performance on the audio dataset. The Evaluation Metrics and the results of these models are described in the later sections.

\begin{table}[h]
\caption{Accuracy, EER and ROC-AUC OF THE BASELINE MODELS on FoR and ASVSPOOF2019 LA TEST SETS}
\begin{center}
\large
\scalebox{0.6}{
\begin{tabular}{|c|c|c|c|c|c|c|c|}
\hline
\textbf{Baseline} & \textbf{Dataset}& \multicolumn{3}{c|}{\textbf{MFCC}} & \multicolumn{3}{c|}{\textbf{LFCC}} \\
\cline{3-8}
\textbf{Model} & \textbf{Name}&\textbf{\textit{Acc(\%)}}& \textbf{\textit{AUC}} & \textbf{\textit{EER}} & \textbf{\textit{Acc(\%)}}& \textbf{\textit{AUC}} & \textbf{\textit{EER}} \\
\hline
RF & FoR (Real+Fake) & 69.05 & 0.785 & 0.309 & 61.175 & 0.682 & 0.374 \\
\hline
MLP & FoR (Real+Fake) & 50.00 & 0.5 & 0.5 & 57.8 & 0.596 & 0.428 \\
\hline
RF & LA (Real+fake) & 45.83 & 0.466 & 0.563 & 52.5 & 0.547 & 0.481 \\
\hline
MLP & LA (Real+fake) & 46.67 & 0.5 & 0.5 & 55.833& 0.564 & 0.461 \\
\hline
\end{tabular}}
\end{center}

\begin{center}
\large
\scalebox{0.6}{%
\begin{tabular}{|c|c|c|c|c|}
\hline
\textbf{End-to-End} & \textbf{\textit{TestSet(\%)}}& \textbf{\textit{Acc(\%)}}& \textbf{\textit{AUC}} & \textbf{\textit{EER}} \\
\hline
TSSDNet & FoR (Real+Fake) & 64.5 & 0.588 & 0.375  \\
\hline
TSSDNet & LA (Real+Fake) &  47.5 & 0.399 & 0.607  \\
\hline
\end{tabular}}
\end{center}
\end{table}

\section{\textbf{Evaluation Metrics}}
To evaluate the performance of the different proposed models on the datasets, we compute the two most commonly used metrics for the evaluation of fake speech detection systems.

\begin{itemize}
\item Equal error rate (EER) is used as the first metric for evaluating the fake speech detection methods. The EER is determined by the point at which the false acceptance rate and the false rejection rate are equal.
\item In order to view the performance of detectors for detecting fake speech from another perspective, we also used the receiver operating characteristic (ROC) curve as a secondary metric. The ROC curve illustrates the relationship between false-positive rate and true-positive rate. We draw the ROC curve and calculate the area under the ROC curve (AUC) for the results of detection. The fake speech detection systems are extremely vulnerable for fake speeches when the ROC points tend towards the diagonal line, and vice versa. 
\end{itemize}

\begin{table}[htbp]
\caption{Accuracy, EER and ROC-AUC OF THE Proposed MODELS on FoR TEST SET}
\begin{center}
\large
\scalebox{0.8}{%
\begin{tabular}{|c|c|c|c|}
\hline
\textbf{ResNet Model} &\textbf{\textit{Acc (\%)}}& \textbf{\textit{EER}} & \textbf{\textit{AUC}}  \\
\hline
MFCC + SEResNet50 & 61.5 & 0.385 & 0.615 \\
\hline
LFCC + SEResNet50 & 50.06 & 0.499 & 0.500 \\
\hline
Masked MFCC + SEResNet50 & 66.36 & 0.337 & 0.663 \\
\hline
Maksed LFCC + SEResNet50 & 50.05 & 0.5 & 0.5 \\
\hline
AFF + SEResNet50 & 71.05 & 0.186 & 0.814 \\
\hline
AFF + SEResNet34 & 95.01 & 0.049 & 0.950 \\
\hline
Masked AFF + SEResNet50 & 79.02 & 0.189 & 0.811 \\
\hline
Masked AFF + SEResNet34 & 90.09 & 0.098 & 0.902 \\
\hline
Pretrained Image Resnet50 & 
49 & - & - \\
\hline
\end{tabular}}
\label{tab1}
\end{center}
\end{table}

\section{\textbf{Results}}
{The Baseline models RF and MLP are trained with FoR train subset and tested with FoR test subset and also on LA dataset to see the performance on the cross-dataset. The pretrained Inc-TSSDNet model was trained on ASVSpoof2019 LA dataset \cite{wang2020asvspoof}, so it was tested using only the FoR test subset. The accuracies and ROC-AUC  values are in the Table-2. The performance results of the Proposed models in Table-3 show that the best model with higher accuracy is AFF+SEResNet34 95\%.} 

The simple machine learning clasisfiers RF and MLP didn't perform well even with the both FoR and LA test sets. So, they definitely can't detect samples from synthetic speech generative model. When carried out experiments with the end-to-end TSSDNet on these synthesizers, it performed comparatively well on some generative models. So, the comparison of the generalizing capacities is made between the end-to-end model and the best of the proposed models. They are evaluated using the samples obtained from different synthesizers like WaveNet \cite{DBLP:journals/corr/OordDZSVGKSK16}, WaveRNN \cite{DBLP:journals/corr/abs-1802-08435}, etc and the results are show in Table-4. We can observe from Table-4 that SE ResNet50 + Masked AFF shows better generalizability compared to the other models.

\begin{table}[htbp]
\caption{Comparing the Generalising power (Acc -\%) of end-to-end model and best of the proposed models}
\begin{center}
\large
\scalebox{0.7}{%
\begin{tabular}{|c|c|c|c|}
\hline
\textbf{Speech} &\textbf{\textit{Inc-TSSDNet}}& \textbf{\textit{SEResNet34}} & \textbf{\textit{SEResNet50}}  \\
\textbf{Synthesizer} &\textbf{\textit{(end-to-end)}}& \textbf{\textit{+ AFF }} & \textbf{\textit{+ MaskedAFF}}  \\
\hline
WaveNet & 100 & 100 & 100 \\
\hline
WaveRNN & 100 & 100 & 100 \\
\hline
FastSpeech & 100 & 80 & 92.31 \\
\hline
Tacotron \& WaveGlow & 88.8 & 78.05 & 94.54 \\
\hline
LA(Fake) & 35 & 87.5 & 98.44 \\
\hline
\end{tabular}}
\label{tab1}
\end{center}
\end{table}

\section{\textbf{Conclusion}}
Through this work we attempted at experimenting feature fusion and application of squeeze excitation to the synthetic speech classification domain. Our results indicate that the models that utilized Attentional Feature Fusion performed well compared to models that were trained using a single feature. It can also be seen that the EER is significantly reduced when compared to the baseline models, including much complex networks like the end-to-end models. Our experiments on the generalizability of these models revealed that ResNets + AFF could also generalize to unseen dataset.

{\footnotesize \bibliographystyle{acm}
\bibliography{ref}}
\vspace{12pt}
\end{document}